\documentclass[12pt,aps,preprint,nofootinbib,superscriptaddress,nobalancelastpage]{revtex4}

\usepackage{hyperref}
\usepackage{epsfig}
\usepackage{amssymb}
\usepackage{amsmath}
\usepackage{amsfonts}
\usepackage{graphics}
\usepackage{cancel}
\usepackage{bbm}
\usepackage{mathrsfs}
\usepackage{slashed}
\usepackage{color}

\newcommand{\mtrx}[2]{\left(\begin{array}{#1} #2 \end{array}\right)}

\newcommand{\Refs}{Refs.}
\newcommand{\Ref}{Ref.}
\newcommand{\Sec}{Sec.}

\newcommand{\Tab}{Tab.}

\newcommand{\eq}{Eq.}

\newcommand{\bea}{\begin{eqnarray}}
\newcommand{\eea}{\end{eqnarray}}

\newcommand{\be}{\begin{equation}}
\newcommand{\ee}{\end{equation}}

\newcommand{\ba}{\begin{array}}
\newcommand{\ea}{\end{array}}

\newcommand{\ie}{\emph{i.e.}}
\newcommand{\eg}{\emph{e.g.}}

\newcommand{\hc}{\mathrm{H.c.}}

\DeclareMathOperator{\diag}{diag}

\allowdisplaybreaks

\begin{document}
\title{Parametrization of Seesaw Models and Light Sterile Neutrinos}

\author{Mattias Blennow}
\email[]{blennow@mppmu.mpg.de}
\affiliation{Max-Planck-Institut f\"ur Physik
(Werner-Heisenberg-Institut), F\"ohringer Ring 6, 80805 M\"unchen,
Germany}

\author{\mbox{Enrique~Fernandez-Martinez}}
\email[]{enfmarti@cern.ch}
\affiliation{CERN Physics Department, Geneve 23, CH-1211 Geneva, Switzerland}

\begin{abstract}

The recent recomputation of the neutrino fluxes from nuclear reactors relaxes the tension between the LSND and MiniBooNE anomalies and disappearance data when interpreted in terms of sterile neutrino oscillations. The simplest extension of the Standard Model with such fermion singlets is the addition of right-handed sterile neutrinos with small Majorana masses. Even when introducing three right-handed neutrinos, this scenario has less free parameters than the 3+2 scenarios studied in the literature. This begs the question whether the best fit regions obtained can be reproduced by this simplest extension of the Standard Model. In order to address this question, we devise an exact parametrization of Standard Model extensions with right-handed neutrinos. Apart from the usual $3 \times 3$ neutrino mixing matrix and the 3 masses of the lightest neutrinos, the extra degrees of freedom are encoded in another $3 \times 3$ unitary matrix and 3 additional mixing angles. The parametrization includes all the correlations among masses and mixings and is valid beyond the usual seesaw approximation. Through this parametrization we find that the best fit regions for the LSND and MiniBooNE anomalies in a 3+2 scenario can indeed be reproduced despite the smaller number of degrees of freedom.

\end{abstract}

\preprint{CERN-PH-TH/2011-179, MPP-2011-89, EURONU-WP6-11-39}

\maketitle

\section{Introduction}

Despite the success of the standard three-neutrino oscillations in explaining the results of solar, atmospheric, reactor and accelerator experiments (see e.g. \Ref~\cite{Schwetz:2011qt}) by two large and one small mixing angles and two distinct mass squared splittings, a number of experiments are now hinting to the existence of oscillations at much shorter baselines. This would imply the existence of extra mass squared splittings $\Delta m^2 \sim 1$~eV and therefore the mixing of the three Standard Model (SM) active neutrinos with extra sterile neutrino states. While, by themselves, none of these hints is very compelling, the fact that they all point towards similar regions of the parameter space ($\Delta m^2 \sim 1$~eV and $|U_{\alpha i}| \sim 0.1$) is intriguing. The better-known and stronger of these anomalies corresponds to the long-standing $3.8 \sigma$ excess of $\bar{\nu}_e$ in the $\bar{\nu}_\mu$ beam observed by the LSND experiment~\cite{Aguilar:2001ty}. This anomaly was explored by the MiniBooNE collaboration and, while it was not confirmed in the neutrino mode~\cite{AguilarArevalo:2007it}, a similar excess, although only at $\sim 2 \sigma$, was found in antineutrinos~\cite{AguilarArevalo:2010wv}. When interpreted as neutrino oscillations, such a difference in the neutrino and antineutrino oscillation probabilities can be explained by the CP violation expected to be present in appearance channels. 

The interpretation of these anomalies as mixing between the active SM neutrinos and new sterile degrees of freedom\footnote{Any new light state mixing with the 3 SM active neutrinos needs to be sterile so as not to contribute the measurement of the invisible width of the $Z$~\cite{Z:2005ema}, which constrains the number of active neutrinos with masses below $M_Z/2$ to 3.} is plagued by the tensions between these positive appearance results and the constraints derived by the negative neutrino disappearance experiments~\cite{Maltoni:2002xd,Sorel:2003hf,Maltoni:2007zf,Karagiorgi:2009nb}, notably reactor experiments~\cite{Kwon:1981ua,Zacek:1986cu,Vidyakin:1987ue,Kuvshinnikov:1990ry,Declais:1994ma,Declais:1994su,Boehm:2001ik,Apollonio:2002gd} for $\bar\nu_e$ and CDHS~\cite{Dydak:1983zq} and atmospheric~\cite{Maltoni:2007zf} neutrinos for $\nu_\mu$ disappearance. This situation has changed recently with the updated computation of the neutrino fluxes from nuclear reactors~\cite{Mueller:2011nm,Huber:2011wv} which predict a slightly larger flux, as compared to the previous estimate, favouring $\bar{\nu}_e$ disappearance in reactors at $2.2 \sigma$ and adding weight to the sterile neutrino interpretation of the LSND and MiniBooNE anomalies~\cite{Mention:2011rk}. Furthermore, the Gallium neutrino anomaly~\cite{Bahcall:1994bq,Giunti:2006bj,Giunti:2010wz,Giunti:2010zu} also seems to favour short baseline $\nu_e$ disappearance. New fits to a 3+2 neutrino oscillation scheme, \ie, 2 extra sterile neutrinos mixing with the 3 SM active ones, show that the tension between appearance and disappearance data is greatly reduced with the new computation of reactor neutrino fluxes, but significant tension remains mainly from the atmospheric and CDHS sector~\cite{Kopp:2011qd,Giunti:2011gz}. Two extra neutrinos are favoured with respect to only one so as to provide the CP violating interference necessary to accommodate the negative results in the MiniBooNE neutrino channel with the excess found in antineutrinos. More experimental results will be required to clarify the present tension between the experiments favouring the existence of short baseline neutrino oscillations and the null result searches.

From a theoretical point of view, it is an interesting exercise to consider the particle content and interaction Lagrangian that could give rise to the 3+2 scenario that currently provides the best (although not-so-good) fit to neutrino data. When these 3+2 mixing scenarios are considered in the literature, a general $5 \times 5$ mixing matrix is normally assumed. Such a matrix can be parametrized by 10 independent mixing angles and 15 phases. Out of the 15 phases, 3 can be reabsorbed in charged lepton field redefinitions, 2 are not physical and 4 are Majorana phases that could be reabsorbed in the neutrino fields, unless they are Majorana particles, and in any case do not play a role in neutrino oscillations. In addition, one of the mixing angles is also unphysical, corresponding to an arbitrariness in the definition of the sterile flavor states. This leaves us with 9 mixing angles and 6 Dirac phases on top of the 5 neutrino masses which are independent and relevant for neutrino oscillations. 

The simplest extension of the SM particle content with singlet fields corresponds to the natural addition of right-handed neutrinos. These right-handed neutrinos should be singlets of the SM gauge group and allow to write Yukawa couplings between the SM active neutrinos and the Higgs field that induce Dirac mass terms after the Higgs develops its vacuum expectation value (vev) and thus account for neutrino masses. Moreover, being SM singlets, right-handed neutrinos are also allowed to have Majorana mass terms. If we consider the addition of 2 right-handed neutrinos, we find that the Yukawa couplings are given by a $3 \times 2$ matrix with 6 independent moduli and the same number of phases. The Majorana masses of the right-handed neutrinos only constitutes 2 extra moduli, since a basis in which the matrix is diagonal and positive can always be chosen. We therefore see that we have only 8 independent moduli and 3 physical phases (one of them Majorana) to be compared with the 14 independent moduli (9 angles and 5 masses) and 6 Dirac phases assumed in 3+2 schemes. Even adding 3 right-handed neutrinos instead of 2, which also seems more natural considering the SM particle content, only brings us up to 12 independent moduli (6 masses and 6 angles) and 4 Dirac phases. Therefore, the question of whether these simple extensions of the SM with a reduced number of degrees of freedom with respect to the general 3+2 scenario can provide the same best fit for neutrino oscillation experiments naturally arises. Notice that, in order to extend the SM by a completely general $5 \times 5$ mixing matrix and 5 independent neutrino masses so as to obtain the 3+2 scenario, a completely general mass matrix mixing the 5 neutrino states (3 active and 2 steriles) would be required. This would imply, for example, to extend the SM with 3 right-handed neutrinos, so as to allow Yukawa couplings and thus Dirac masses plus another extra 4 SM singlets, 2 left-handed and 2 right-handed, forming 2 Dirac pairs with masses unrelated to the Higgs mechanism and thus providing a general $5 \times 5$ mass matrix. This particle content seems highly unnatural compared to the simple addition of 3 light right-handed Majorana neutrinos.  

Motivated by this question, we have devised a new parametrization of models with extra right-handed neutrinos as, for example, the popular type-I seesaw mechanism~\cite{Minkowski:1977sc,Mohapatra:1979ia,Yanagida:1979as,GellMann:1980vs}. This parametrization takes into account all the relations between masses and mixing parameters implied by the reduced number of degrees of freedom discussed above. While several such parametrizations already exist in the literature, we find that they have been derived in the seesaw limit, \ie, for Majorana masses much larger than the Dirac ones. While this is usually an excellent approximation in seesaw models, we found it unsuited for the task at hand, since the masses of the new sterile states required for the LSND and MiniBooNE anomalies are around 1~eV, not far from the constrained range of the 3 SM neutrinos. Other variants of the seesaw mechanism, such as the linear~\cite{Malinsky:2005bi} or the inverse~\cite{Mohapatra:1986bd} seesaws, also have very similar Dirac and Majorana masses and thus the standard seesaw approximation and the parametrizations derived from it might not be suitable. The new parametrization we introduce is exact and makes no assumption on the sizes of the Majorana and Dirac mass matrices. In the case in which 3 extra right-handed neutrinos are introduced, the 12 independent moduli and 6 phases are distributed in a $3 \times 3$ unitary matrix, which reduces to the standard one when the mixing between the sterile and active neutrinos is small, the 3 masses of the lightest neutrino mass eigenstates, 3 mixing angles which are related to the 3 masses of the heavier mass eigenstates and an extra $3 \times 3$ unitary matrix, making a total of 3 masses, 9 angles and 6 phases (4 Dirac and 2 Majorana).   

In \Sec~\ref{sec:param} we introduce the new parametrization comparing it with other existing parametrizations of the seesaw limit in the literature and discussing the physical ranges of the parameters introduced and the relations between them. In \Sec~\ref{sec:examples} we apply the parametrization to three widely different examples: the seesaw limit, the case of purely Dirac neutrinos and the intermediate regime of light Majorana masses that can provide the best fit to the LSND and MiniBooNE anomalies, explicitly showing that the 3+2 best fit can be reproduced despite the reduced number of degrees of freedom. Finally, in \Sec~\ref{sec:summary} we summarize the results and draw our conclusions.

\section{Parametrization}
\label{sec:param}

Here we discuss a new parametrization of the unknown degrees of freedom of the neutrino sector necessary to fully specify the Lagrangian and physical observables when extending the SM by fermion singlets (right-handed neutrinos) as in the type-I seesaw mechanism (but with arbitrary masses). For simplicity we will here consider the case in which three extra right-handed neutrinos are added $N_\mathrm{R}^i$. The Lagrangian would thus read:
\begin{eqnarray}\label{eq:The3FormsOfNuMassOp}
\mathscr{L} &=& \mathscr{L}_\mathrm{SM} -\frac{1}{2} \overline{N_\mathrm{R}^i} (M_{N})_{ij} N^{c j}_\mathrm{R} -(Y_{N})_{i\alpha}\overline{N_\mathrm{R}^i}  \phi^\dagger
\ell^\alpha_\mathrm{L} +\hc\; .
\end{eqnarray}
Here, $\phi$ denotes the SM Higgs field, which breaks the electroweak (EW) symmetry after acquiring its vev $v_{\mathrm{EW}}$. We have also introduced the Majorana mass allowed for the right-handed neutrinos $M_N$ as well as the Yukawa couplings between the neutrinos and the Higgs field. The vev of the Higgs will also induce Dirac masses $m_D = v_{EW} Y_N$. Thus, the full $6\times 6$ mixing matrix $U_{\rm tot}$ is the unitary matrix that diagonalizes the extended neutrino mass matrix: %
\begin{equation}
U_{\rm tot}^T \left(
\begin{array}{cc}
0 & m_D^T \\ m_D & M_N
\end{array}
\right) U_{\rm tot} = \left(
\begin{array}{cc}
m & 0 \\ 0 & M
\end{array}
\right), \label{eq:diag}
\end{equation}
where $m$ and $M$ are diagonal matrices. Without loss of generality this diagonalization can be performed in two steps: first
a block-diagonalization and then two unitary rotations to diagonalize the mass matrices of the light and heavy neutrinos, \ie, %
\begin{equation}
U_{\rm tot} = \left(
\begin{array}{cc}
A_{11} & A_{12} \\ A_{21} & A_{22}
\end{array}
\right) \left(
\begin{array}{cc}
U & 0 \\ 0 & U'
\end{array}
\right), \label{eq:block}
\end{equation}
where $U$ and $U'$ are unitary matrices. We can always choose a basis for the heavy singlets such that $U'=I$ (rotations amongst the sterile states are unphysical). The two steps of the unitary rotation can be expressed as the exponential of antihermitian matrices, which correspond to the Lie algebra of the unitary group. In particular the block-diagonalization will correspond to the exponential of a block-off diagonal antihermitian matrix:
\begin{equation}
\left(
\begin{array}{cc}
A_{11} & A_{12} \\ A_{21} & A_{22}
\end{array}
\right) = \exp \left(
\begin{array}{cc}
0 & \Theta \\ -\Theta^\dagger & 0
\end{array}
\right) = \left(
\begin{array}{cc}
\displaystyle\sum\limits_{n=0}^\infty \frac{ \left(- \Theta \Theta^\dagger \right)^{n}}{(2n)!} & 
\displaystyle\sum\limits_{n=0}^\infty \frac{ \left(- \Theta \Theta^\dagger \right)^{n}}{\left(2n+1\right)!} \Theta  \\ 
-\displaystyle\sum\limits_{n=0}^\infty \frac{ \left(- \Theta^\dagger \Theta \right)^{n}}{\left(2n+1\right)!} \Theta^\dagger & 
\displaystyle\sum\limits_{n=0}^\infty \frac{ \left(- \Theta^\dagger \Theta \right)^{n}}{(2n)!}
\end{array}
\right) ,
\label{eq:exp}
\end{equation}
With $\Theta$ a completely general $3 \times 3$ matrix. The series of \eq~(\ref{eq:block}) correspond to somewhat modified versions of the sine and cosine series, we therefore define: 
\begin{equation}
\left(
\begin{array}{cc}
\ c & s \\ -s^\dagger & \hat{c}
\end{array}
\right) \equiv \left(
\begin{array}{cc}
\displaystyle\sum\limits_{n=0}^\infty \frac{ \left(- \Theta \Theta^\dagger \right)^{n}}{(2n)!} & 
\displaystyle\sum\limits_{n=0}^\infty \frac{ \left(- \Theta \Theta^\dagger \right)^{n}}{\left(2n+1\right)!} \Theta  \\ 
-\displaystyle\sum\limits_{n=0}^\infty \frac{ \left(- \Theta^\dagger \Theta \right)^{n}}{\left(2n+1\right)!} \Theta^\dagger & 
\displaystyle\sum\limits_{n=0}^\infty \frac{ \left(- \Theta^\dagger \Theta \right)^{n}}{2n!}\end{array}
\right).
\label{eq:sincos}
\end{equation}
Substituting this expression into \eq~(\ref{eq:diag}) we obtain: 
\begin{equation}
\label{eq:corr}
c^* U^* m U^\dagger c = - s^* M s^\dagger .
\end{equation}
This relation contains all the correlations between the active neutrino masses and mixings and the sterile ones and is the starting point of any parametrization. In the seesaw limit $M_N \gg m_D$ so that
\begin{equation}
\Theta \simeq m_D^\dagger M_N^{-1}
 \label{eq:theta}
\end{equation}
represents the generalized $3 \times 3$ mixing between the active neutrinos and the heavy mass eigenstates and $U$ corresponds to the $3 \times 3$ low energy neutrino mixing matrix between the flavour states with high accuracy. Then \eq~(\ref{eq:corr}) simplifies to the well-known relation:
\begin{equation}
\label{eq:corrseesaw}
U^* m U^\dagger  = - m_D^t M^{-1} m_D .
\end{equation}
This last equation has been used to introduce several seesaw parametrizations. 
For example, the popular Casas-Ibarra parametrization involving an orthogonal complex matrix $R$ introduced in \cite{Casas:2001sr} exploits the fact that from \eq~(\ref{eq:corrseesaw}) the matrix $R = i M^{-1/2} m_D U m^{-1/2}$ has to be orthogonal and thus the neutrino Dirac masses can be easily obtained in terms the mass eigenstates $m$ and $M$, the low energy mixing matrix $U$ and the elements of the matrix $R$ as $m_D = -i M^{1/2} R m^{1/2} U^\dagger$. The main advantage of this parametrization is that the heavy mass eigenstates in $M$ are part of its set of free parameters. However, the physical range of the parameters contained in $R$ can be cumbersome and lead to complications in the scan of the seesaw parameter space. Indeed, $R$ can be parametrized by three complex angles with unconstrained imaginary parts, see \Ref~\cite{Casas:2010wm} for a detailed discussion. 

Another convenient and popular parametrization, discussed in detail for example in \cite{Casas:2006hf}, is obtained by diagonalizing the Dirac mass matrix through two unitary rotations: $m_D = U_R D U_L^\dagger$ where $D$ is a real diagonal matrix and $U_R$ and $U_L$ are unitary. Substituting in \eq~(\ref{eq:corrseesaw}): 
\begin{equation}
\label{eq:CD}
D^{-1} U^t_L U^* m U^t U_L D^{-1} = - U_R^t M^{-1} U_R .
\end{equation}
Thus, the heavy mass eigenstates $M$ and the mixing $U_R$ can be parametrized in terms of the light masses and mixings $m$ and $U$ together with a unitary matrix $U_L$ and the three real numbers contained in $D$ by performing the diagonalization of \eq~(\ref{eq:CD}). This parametrization replaces the complex orthonormal matrix of the Casas-Ibarra parametrization, $R$, by a unitary matrix, $U_L$, allowing more convenient scans of the parameter space. Conversely, the heavy mass eigenstates are no longer free parameters but derived. 

A generalization of this second parametrization away from the seesaw in which $M_N \gg m_D$ limit was performed in \Ref~\cite{Donini:2011jh}. In that case, the Lagrangian parameters, \ie, $D$, $U_R$, $U_L$ and $M_N$ where used as free parameters. While this choice allows a very simple reconstruction of the Lagrangian, relating these parameters with the mass eigenvalues and mixing angles can be cumbersome and we therefore propose a different approach.

\subsection{Parametrization outside the seesaw limit}

In order to make an exact parametrization, we start with \eq~(\ref{eq:corr}) and introduce a biunitary diagonalization of $\Theta$:
\begin{equation}
\Theta = V_L \theta V^\dagger_R ,
\label{eq:thetadiag}
\end{equation}
where $\theta$ is a positive diagonal matrix of angles. Under this diagonalization $s = V_L \sin(\theta) V^\dagger_R$, $c = V_L \cos(\theta) V^\dagger_L$ and $\hat{c} = V_R \cos(\theta) V^\dagger_R$. Thus, \eq~(\ref{eq:corr}) becomes: 
\begin{equation}
\cot(\theta) V^t_L U^* m U^\dagger V_L \cot(\theta) = - V_R^t M V_R.
\label{eq:massrel}
\end{equation}
Therefore, the masses $M$ and the mixing $V_R$ can be obtained in terms of the masses $m$ and mixings $U$, together with a unitary matrix $V_L$ and the three real angles contained in $\theta$ by performing a diagonalization. The full mixing matrix is then given by:
\begin{equation}
U_{\rm tot} = \left(
\begin{array}{cc}
V_L \cos(\theta) V^\dagger_L U & V_L \sin(\theta) V^\dagger_R \\ -V_R \sin(\theta) V^\dagger_L U & V_R \cos(\theta) V^\dagger_R
\end{array}
\right). \label{eq:result}
\end{equation}
Regarding the independent phases contained in the matrices $V_L$ and $U$, parameter counting tells us that the original Lagrangian had 9 independent phases, since all the elements of the Dirac mass matrix $m_D$ can be complex. Of these 9 phases, 3 can be absorbed by field redefinitions of the charged leptons. This is not an option for the neutrino fields since they have Majorana mass terms. To identify the remaining 6 independent phases we write down the three unitary matrices in the form $\Phi_1 V \Phi_2$, where $\Phi_1$ and $\Phi_2$ are diagonal matrices of phases and $V$ has the usual CKM form. 
\begin{equation}
V(\theta_{12},\theta_{23},\theta_{13},\delta) = \left(
\begin{array}{ccc}
1 & 0 & 0 \\ 0 & c_{23} & s_{23} \\ 0 & -s_{23} & c_{23}
\end{array}
\right)
\left(
\begin{array}{ccc}
c_{13} & 0 & s_{13} e^{-i\delta} \\ 0 & 1 & 0 \\ -s_{13} e^{i\delta} & 0 & c_{13}
\end{array}
\right)
\left(
\begin{array}{ccc}
c_{12} & s_{12} & 0 \\ -s_{12} & c_{12} & 0 \\ 0 & 0 & 1
\end{array}
\right). \label{eq:CKM}
\end{equation}
Notice that, from the definition of $V_L$ and $V_R$ in \eq~(\ref{eq:thetadiag}), the phases contained in $\Phi_2$ for the matrix $V_L$ can be absorbed in the definition of $V_R$. As for the matrix $U$, the phases contained in $\Phi_2$ are the usual Majorana phases of the $3 \times 3$ neutrino mixing matrix, while those contained in $\Phi_1$ either appear in combination with the $\Phi_1$ of $V_L$ or can be absorbed via the field redefinitions of the charged leptons. Therefore, only the $\Phi_1$ of either $U$ or $V_L$ is independent. We then choose the following parametrization for the two independent unitary matrices $U = V(\theta_{12},\theta_{23},\theta_{13},\delta) \cdot \diag(1, e^{-i\alpha_2},e^{-i\alpha_3})$ and $V_L = \diag(1, e^{-i\alpha^L_2},e^{-i\alpha^L_3}) \cdot V(\theta^L_{12},\theta^L_{23},\theta^L_{13},\delta^L)$. Notice that $\alpha_2$ and $\alpha_3$ can be considered Majorana phases, since they would become unphysical if the active neutrinos were Dirac fields, while $\delta$, $\delta^L$, $\alpha^L_2$ and $\alpha^L_3$ correspond to Dirac phases.

An alternative approach to the parametrization suggested above is to perform the equivalent of the Casas-Ibarra approach outside of the seesaw limit. In this case, we introduce the matrix
\begin{equation}
 B = i \sqrt{M^{-1}} s^{-1} c U \sqrt{m},
\end{equation}
which by virtue of \eq~(\ref{eq:corr}) is complex orthogonal. From this definition we can do the computation
\begin{equation}
 F \equiv i U \sqrt{m} B^T \sqrt{M^{-1}} = c^{-1} s = V_L \tan(\theta) V_R^\dagger,
\end{equation}
where $U$, $M$, $m$ and $B$ are choosen as the independent set of parameters, while $V_L$, $V_R$ and $\theta$ are quantities that can be computed from the bidiagonalization of $F$. Once all the quantities are known, the full mixing matrix $U_{\rm tot}$, and therefore also the fundamental parameters in the Lagrangian, can be computed using \eq~(\ref{eq:result}). The advantage of this parametrization is that it contains all of the physical masses as parameters, while the disavantage is that it does not have a clear interpretation of the complex orthogonal matrix $B$ (while in our first parametrization, $\theta$ is directly related to the active-sterile mixing).

In what follows we will use the first parametrization, whose independent parameters consists of $m$, $U$, $V_L$, and $\theta$.

\subsection{Physical ranges}

We will now discuss the physical ranges that the angles and phases that define the neutrino mixing matrix should take in order to reproduce all the distinct physical situations avoiding overcounting. Notice that, apart from the freedom that allows us to redefine the charged lepton fields up to an overall phase used to absorb three of the phases, in the neutrino fields we can also perform the transformation $\nu_i \to -\nu_i$ and their Majorana masses remain unchanged. Therefore, following \Ref~\cite{deGouvea:2008nm}, two sets of mixing parameters $\theta_{i,j}$, $\delta$, $\alpha_i$, $\theta^L_{i,j}$, $\delta^L$, $\alpha^L_i$ and $\theta'_{i,j}$, $\delta'$, $\alpha'_i$, $\theta'^L_{i,j}$, $\delta'^L$, $\alpha'^L_i$ will describe the same physical situation if the corresponding mixing matrices are related through the transformations:
\begin{equation}
U_{\rm tot}(\theta'_{i,j},\delta',\alpha'_i,\theta'^L_{i,j},\delta'^L,\alpha'^L_i) =  P^l U_{\rm tot}(\theta_{i,j},\delta,\alpha_i,\theta^L_{i,j},\delta^L,\alpha^L_i) P^n,
\label{eq:transf}
\end{equation}
where $P^l$ and $P^n$ are diagonal matrices with either -1 or 1 in the diagonal and represent the corresponding field redefinitions of charged leptons and neutrinos respectively. Moreover, only the first three rows of $U_{\rm tot}$ in \eq~(\ref{eq:result}) are physical, since only the active neutrinos take part in charged current interactions. Therefore, \eq~(\ref{eq:transf}) reduces to:
\begin{eqnarray}
\nonumber
V'_L \cos(\theta) V'^\dagger_L U' = P^l V_L P  \cos(\theta) P V_L^\dagger P^l P^l U P^n \\
V'_L \sin(\theta) V'^\dagger_R = P^l V_L P \sin(\theta) P V_R^\dagger P^s, 
\label{eq:transf2}
\end{eqnarray}
where, for simplicity, we have omitted the explicit dependence of the matrices on the angles and phases defining $V'_L = V_L(\theta'^L_{i,j},\delta'^L,\alpha'^L_i)$ and so on. $P$ and $P^s$ are arbitrary matrices with the same structure as $P^l$ and $P^n$. Thus, transformations of the form 
\begin{eqnarray}
\nonumber
V'_L  = P^l V_L P  \\
\nonumber
V'_R = P^s V_R P \\
U' = P^l U P^n,
\label{eq:transf3}
\end{eqnarray}
represent the same physical situation. Since $V_R$ is not independent from $V_L$ and $U$ it is important to confirm the consistency of the set of transformations of \eq~(\ref{eq:transf3}) with \eq~(\ref{eq:massrel}). 

\begin{table}
\begin{center}
\begin{tabular}{|c|c|c|c|c|}
\hline
$P^l_{ii}$ & $P^n_{ii}$ & $P_{ii}$ & $\theta'_{i,j},\delta',\alpha'_i$, $\theta'^L_{i,j},\delta'^L,\alpha'^L_i$ & Restricts \\
\hline
\hline
1,1,1 & 1,1,1 & 1,1,1 & 
$\theta_{12}+\pi$, $\theta_{13}+\pi$, $\theta_{23}+\pi$, $\delta+\pi$ &
$\theta_{13} \in [-\pi/2,\pi/2]$ \\
\hline
1,1,1 & 1,1,1 & 1,1,1 & 
$\theta^L_{12}+\pi$, $\theta^L_{13}+\pi$, $\theta^L_{23}+\pi$, $\delta^L+\pi$ &
$\theta^L_{13} \in [-\pi/2,\pi/2]$ \\
\hline
1,-1,-1 & 1,1,1 & 1,1,1 & 
$\theta_{23}+\pi$, $\alpha^L_{2}+\pi$, $\alpha^L_{3}+\pi$ & 
$\theta_{23} \in [-\pi/2,\pi/2]$ \\
\hline
1,1,1 & 1,1,1 & 1,1,1 & 
$\theta^L_{23}+\pi$, $\alpha^L_{2}+\pi$, $\alpha^L_{3}+\pi$ & 
$\theta^L_{23} \in [-\pi/2,\pi/2]$ \\
\hline
-1,-1,-1 & 1,1,-1 & -1,-1,-1 & 
$\theta_{12}+\pi$ &
$\theta_{12} \in [-\pi/2,\pi/2]$ \\
\hline
1,1,1 & 1,1,1 & -1,-1,1 & 
$\theta^L_{12}+\pi$ & 
$\theta^L_{12} \in [-\pi/2,\pi/2]$ \\
\hline
1,-1,1 & 1,-1,1 & 1,-1,1 & 
$-\theta_{12}$, $-\theta_{23}$, $-\theta^L_{12}$, $-\theta^L_{23}$ & 
$\theta^L_{23} \in [0,\pi/2]$ \\
\hline
1,-1,1 & 1,-1,1 & 1,1,1 & 
$-\theta_{12}$, $-\theta_{23}$, $\alpha^L_{2}+\pi$ &
$\theta_{23} \in [0,\pi/2]$ \\
\hline
1,-1,-1 & 1,-1,-1 & 1,1,1 & 
$-\theta_{12}$, $-\theta_{13}$, $\alpha^L_{2}+\pi$, $\alpha^L_{3}+\pi$ &
$\theta_{12} \in [0,\pi/2]$ \\
\hline
1,1,1 & 1,1,1 & 1,-1,-1 & 
$-\theta^L_{12}$, $-\theta^L_{13}$, $\alpha^L_{2}+\pi$, $\alpha^L_{3}+\pi$ &
$\theta^L_{12} \in [0,\pi/2]$ \\
\hline
1,1,1 & 1,1,1 & 1,1,1 & 
$-\theta_{13}$, $\delta+\pi$ &
$\theta_{13} \in [0,\pi/2]$ \\
\hline
1,1,1 & 1,1,1 & 1,1,1 & 
$-\theta^L_{13}$, $\delta^L+\pi$ &
$\theta^L_{13} \in [0,\pi/2]$ \\
\hline
1,1,1 & 1,-1,1 & 1,1,1 & 
$\alpha_2+\pi$ &
$\alpha_2 \in (-\pi/2,\pi/2]$ \\
\hline
1,1,1 & 1,1,-1 & 1,1,1 & 
$\alpha_3+\pi$ &
$\alpha_3 \in (-\pi/2,\pi/2]$ \\
\hline
\end{tabular}
\caption{Set of transformations of angles and phases that describe the same physical configuration. The restrictions that each transformation allows to set on the phases and mixing angles in order to bring them to their physical range is also displayed.}
\label{tab:phys}
\end{center}
\end{table}

Following \Ref~\cite{deGouvea:2008nm} we show that with different choices of $P^l$, $P^n$ and $P$ a set of transformations among phases and angles can be defined such that the physical range of the 6 angles can be chosen to be $[0,\pi/2]$, of the 2 Majorana phases $(-\pi/2,\pi/2]$ and $(-\pi,\pi]$ for the 4 Dirac phases. Table~\ref{tab:phys} shows the list of transformations among the angles and phases that can be used to bring them to their physical ranges and the corresponding values that $P^l$, $P^n$ and $P$ take for each of them. Notice that when $P^l$ or $P^n$ transform, the corresponding charged lepton or neutrino field redefinitions are required to recover the original configuration.

\section{Applied examples}
\label{sec:examples}

In order to get a more intuitive handle on the proposed parametrization, we will now examine how it applies to different known models. In particular, we will discuss the limiting cases of the type-I seesaw limit and Dirac neutrinos, as well as the intermediate regime of light Majorana masses that can provide the best fit to the LSND and MiniBooNE anomalies, explicitly showing that the 3+2 best fit can be reproduced despite the reduced number of degrees of freedom.

\subsection{The seesaw limit}

In the seesaw limit $M_N \gg m_D$ we can expand $\Theta$ to first order and obtain \eq~(\ref{eq:theta}). In addition, the basis where $U' = I$ corresponds to the basis where $M_N$ is already diagonal also to first order, such that $\Theta \simeq m_D^\dagger M^{-1}$. Furthermore, since $\Theta$ is small, we have $c \simeq 1$ and $s \simeq \Theta$. Using this last relation, it is easy to deduce
\begin{equation}
 V_L \sin^2(\theta) V_L^\dagger \simeq V_L \theta^2 V_L^\dagger = \Theta \Theta^\dagger = m_D^\dagger M^{-2} m_D.
\end{equation}
The last equality corresponds to the coefficient of the $d=6$ operator that is obtained in the type-I seesaw after integrating out the heavy neutrino degrees of freedom~\cite{Broncano:2002rw}. Thus, in this limit, the extra degrees of freedom $\theta$ and $V_L$ are simply obtained from the diagonalization of the $d=6$ operator. The remaining parameters, \ie, $m$ and $U$ are contained, as expected in the coefficient of the Weinberg $d=5$ operator from \eq~(\ref{eq:corrseesaw}), responsible for the light neutrino masses. Thus, we find that, as expected, the $d=5$ and $d=6$ operator suffice to reproduce all the parameters of the original Lagrangian, as well as the full mass and mixing matrices~\cite{Broncano:2002rw}.

Comparing with the $U_L$ parametrization of \Ref~\cite{Casas:2006hf}, $D = U_R^\dagger m_D U_L$, we have
\begin{equation}
 M^{-1} m_D = M^{-1} U_R D U_L^\dagger = \Theta^\dagger \simeq V_R \theta V_L^\dagger.
\end{equation}
Thus, the $\theta$ of our parametrization is intimately tied to $D$ through this relation, and so are $V_L$ and $V_R$ with $U_L$ and $U_R$, with the difference that, in this Seesaw limit, our parametrization corresponds to bidiagonalizing $M^{-1} m_D$ rather than $m_D$. It is also easy to relate $\theta$ to the Casas-Ibarra parametrization through the $R$ matrix in a similar fashion, expressing $m_D$, and thus $\Theta$, as $\Theta = -i M^{-1/2} R m^{1/2} U^\dagger$. The procedure to obtain the matrix $R$ from the $d=6$ operator, \ie{} from $\theta$ and $V_L$, can be found in \Refs~\cite{Broncano:2003fq,Antusch:2009gn}.

\subsection{Dirac neutrinos}

In the case when $M_N = 0$, we recover the limit of Dirac neutrinos. In this limit, the full mass matrix has the form
\begin{equation}
 \mathcal M_D = \mtrx{cc}{0 & m_D^T \\ m_D & 0}
\end{equation}
and can be diagonalized as
\begin{equation}
 U_{\rm tot}^T \mathcal M_D U_{\rm tot} = \mtrx{cc}{ - m & 0 \\ 0 & m},
\end{equation}
where
\begin{equation}
 U_{\rm tot} = \frac 1{\sqrt 2}\mtrx{cc}{V_L^D & V_L^D \\ -V_R^D & V_R^D}
\end{equation}
and $m = V_L^{DT} m_D^T V_R^D$ is diagonal. However, $V_R^D$ is an arbitrary rotation of the sterile right-handed component of the Dirac neutrinos and we can therefore choose a basis where $V_R^D = 1$.

In the parametrization we propose, this is reproduced in the case where $\theta = \pi/4$, which corresponds to maximal mixing between the laft and righ-handed fields, as expected. In this case, \eq~(\ref{eq:massrel}) turns into
\begin{equation}
 V_L^T U^* m U^\dagger V_L = -V_R^T M V_R,
\end{equation}
with the obvious solution $V_R = U^\dagger V_L$ and $M = -m$. Inserting this into \eq~(\ref{eq:thetadiag}) for $U_{\rm tot}$, we obtain
\begin{equation}
 U_{\rm tot} = \frac{1}{\sqrt 2}\mtrx{cc}{U & V_L V_R^\dagger \\ -V_R V_L^\dagger U & 1} =
 \frac{1}{\sqrt 2} \mtrx{cc}{U & U \\ -1 & 1},
\end{equation}
which exactly corresponds to the scenario of Dirac neutrinos. It should be noted that there is an arbitrariness in $U = V_L V_R^\dagger$, since any combination of $V_L$ and $V_R$ satisfying this condition is equally valid. This is related to the fact that the rotations $V_R^D$ of the right-handed neutrino fields are unphysical and thus only the physical mixing matrix $U$ remains.

\subsection{LSND/MiniBooNE 3+2 best fit}

We will now address the question of whether the best fit found in the 3+2 scheme with a general $5 \times 5$ mixing matrix for the short baseline neutrino oscillation experiments can be reproduced with the addition of only 3 right-handed neutrinos to the SM particle content. Notice that, while the general 3+2 scheme has a $5 \times 5$ mixing matrix with 9 independent mixing angles, 6 Dirac phases and 5 masses that can play a role in neutrino oscillations, the only relevant parameters are the 4 mass square splittings, the elements of the $3 \times 3$ submatrix relevant for solar and atmospheric neutrino oscillations, and 4 moduli and a phase for the short baseline oscillations that could accommodate LSND and MiniBooNE. Two studies have been performed so far in the 3+2 framework including the new reactor fluxes, obtaining slightly different results for the best fit values of the extra mass splitting and elements of the mixing matrix~\cite{Kopp:2011qd,Giunti:2011gz}. The respective best fits are reported in \Tab~\ref{tab:sterilefit}. Notice that, apart from the two extra splittings $\Delta m^2_{41}$ and $\Delta m^2_{51}$, only the moduli $|U_{e4}|$, $|U_{\mu4}|$, $|U_{e5}|$ and $|U_{\mu5}|$ are relevant for short baseline neutrino oscillations together with the CP violating phase $\phi = \arg(U_{e4} U_{\mu4}^* U_{e5}^* U_{\mu5})$ that can accommodate the null results of MiniBooNE in neutrinos and the positive signal in antineutrinos.

\begin{table}
\begin{center}
\begin{tabular}{|c|c|c|c|c|c|c|c|}
\hline
 & $|U_{e 4}|$ & $|U_{e 5}|$ & $|U_{\mu 4}|$ & $|U_{\mu 5}|$ & $\phi/\pi$  & $\Delta m^2_{41}$ & $\Delta m^2_{51}$  \\
\hline
\hline
\Ref~\cite{Kopp:2011qd} &  0.128 & 0.138 & 0.165 & 0.148 & 1.64 & 0.47 & 0.87 \\
\hline
NH1  &  0.126 & 0.131 & 0.163 & 0.142 & 1.64 & 0.47 & 0.88 \\
\hline
IH1  &  0.118 & 0.138 & 0.160 & 0.156 & 1.64 & 0.47 & 0.88 \\
\hline
\hline
\Ref~\cite{Giunti:2011gz} &  0.130 & 0.130 & 0.134 & 0.080 & 1.52 & 0.9 & 1.6 \\
\hline
NH2 &  0.128 & 0.135 & 0.134 & 0.080 & 1.52 & 0.9 & 1.6 \\
\hline
IH2 &  0.130 & 0.130 & 0.137 & 0.080 & 1.52 & 0.9 & 1.6 \\
\hline
\end{tabular}
\caption{Best fit points for the 3+2 scenario to short baseline neutrino oscillation data performed in \Refs~\cite{Kopp:2011qd,Giunti:2011gz} compared with some sample points chosen to reproduce these values with either a normal or inverted hierarchy pattern. The values of the parameters for these sample points are shown in \Tab~\ref{tab:paramfit}. $\Delta m^2_{41}$ and $\Delta m^2_{51}$ are given in eV$^2$.}
\label{tab:sterilefit}
\end{center}
\end{table}

The possibility of explaining the LSND anomaly through the inclusion of light right-handed Majorana neutrinos was discussed in \Ref~\cite{deGouvea:2006gz}, while the first steps towards answering the question of whether the best fit of \Ref~\cite{Kopp:2011qd} could be reproduced by adding 2 right-handed neutrinos to the SM Lagrangian were already given in \Ref~\cite{Donini:2011jh}. There it was shown that, when assuming all mixing elements real, the values of $|U_{e4}|$, $|U_{\mu4}|$, $|U_{e5}|$ and $|U_{\mu5}|$ were in the correct ballpark compared to those required for short baseline neutrino oscillations when obtaining the correct pattern of masses and mixings in the solar and atmospherics sectors with an inverted hierarchy pattern. Therefore, it is expected that, when extending the analysis to the complex case introducing phases, the best fit would be possible to obtain. 

Here we will extend the analysis to the addition of 3 right handed neutrinos and show that both the best fits from \Refs~\cite{Kopp:2011qd,Giunti:2011gz} can be obtained through our parametrization while at the same time reproducing the correct solar and atmospheric masses and mixings pattern with either a normal or an inverted hierarchy. We have performed a scan of the parameter space through Markov Chain Monte Carlo techniques (MCMC) through the MonteCUBES software~\cite{Blennow:2009pk} and selected 4 points that reproduce well the best fits from \Refs~\cite{Kopp:2011qd,Giunti:2011gz} for either a normal or an inverted hierarchy. The values for the relevant parameters for short baseline neutrino oscillations are shown in \Tab~\ref{tab:sterilefit} and compared to the best fit values from \Refs~\cite{Kopp:2011qd,Giunti:2011gz}. The points are labelled either with NH or IH depending on the hierarchy assumed and 1 or 2 when trying to reproduce the best fit from \Ref~\cite{Kopp:2011qd} or \Ref~\cite{Giunti:2011gz} respectively. The elements of the $U$ matrix (except for the phases) were fixed to the current best fit points of solar and atmospheric neutrino oscillations from Ref~\cite{Schwetz:2011qt}, since we found that $3 \times 3$ mixing matrix that governs these oscillations, \ie{} $V_L^\dagger \cos(\theta) V_L U$, did not deviate much from $U$ and we always obtained its elements within the $1 \sigma$ region. The $\Delta m^2_{21}$ and $\Delta m^2_{31}$ splitting were similarly fixed to their best fit values assuming either a normal or an inverted hierarchy. We also fixed $\theta_{13}=0.1$, a relative large value motivated by the recent hint of sizable $\theta_{13}$ from the T2K experiment~\cite{Abe:2011sj}. The rest of the parameters were kept free in the MCMC and their values are shown in \Tab~\ref{tab:paramfit} for each of the 4 points chosen.  

\begin{table}
\begin{center}
\begin{tabular}{|c|c|c|c|c|}
\hline
 & NH1 & IH1 & NH2 & IH2 \\
\hline
$m_1$ (eV), $\delta$, $\delta^L$ & 0.02, $38^\circ$, $126^\circ$ & 0.01, $-132^\circ$, $132^\circ$ & 0.02, $12^\circ$, $-168^\circ$ & 0.02, $160^\circ$, $166^\circ$  \\
\hline
$\alpha_2$, $\alpha_3$ & $-16^\circ$, $42^\circ$ & $-88^\circ$, $-4.9^\circ$ & $39^\circ$, $34^\circ$ & $84^\circ$, $38^\circ$  \\
\hline
$\theta_{12}^L$, $\theta_{23}^L$, $\theta_{13}^L$ & $78^\circ$, $52^\circ$, $45^\circ$ & $58^\circ$, $47^\circ$, $56^\circ$ & $61^\circ$, $27^\circ$, $27^\circ$ & $64^\circ$, $33^\circ$, $55^\circ$  \\
\hline
$\alpha_2^L$, $\alpha_3^L$ & $-120^\circ$, $-118^\circ$ & $-45^\circ$, $147^\circ$ & $-77^\circ$, $-74^\circ$ & $-46^\circ$, $-82^\circ$  \\
\hline
$\theta_{1}$, $\theta_{2}$, $\theta_{3}$ & $9.8^\circ$, $18^\circ$, $1.6^\circ$ & $12^\circ$, $17^\circ$, $1.3^\circ$ & $6.2^\circ$, $13^\circ$, $3.0^\circ$ & $11^\circ$, $13^\circ$, $0.31^\circ$  \\
\hline
\end{tabular}
\caption{Values of the free parameters in the MCMC scan corresponding to the 4 sample points chosen. The mixing angles of the $U$ matrix as well as $m_2$ and $m_3$ were fixed to their present best fit value for solar and atmospheric neutrino oscillations taken from \Ref~\cite{Schwetz:2011qt}.}
\label{tab:paramfit}
\end{center}
\end{table}

The value of the third mass splitting $\Delta m^2_{61}$ is controlled by the value of the angle $\theta_3$. We chose a small value in all cases that tends to make this third neutrino heavy and its mixings small, so that it mostly decouples. In the examples displayed it was close to the keV scale. However, we found that changing this parameter does not affect much the rest of the values. Thus, decreasing $\theta_3$, the third neutrino can be made much heavier and make it decouple while still keeping good agreement with the best fits of \Refs~\cite{Kopp:2011qd,Giunti:2011gz}. One possibility is to choose $\theta_3 \sim 0.1^\circ-1^\circ$. In this case, the third neutrino will have a mass around a few keV and mixings of order $10^{-4}-10^{-3}$ with the active neutrinos, close to the region in which dark matter could be explained by sterile neutrinos (see, \eg, \Ref~\cite{Boyarsky:2009ix}). On the other hand, decreasing $\theta_3$ to smaller values increases the mass of the sixth neutrino as $1/(\theta_3)^2$ and reduces its couplings to the active ones as $\theta_3$,  making it decouple. Notice, however, that its contribution to light neutrino masses, \ie{} to the Weimberg $d=5$ operator, does not decouple in this manner, since the combination $m_{ab} = U^*_{a6} M_6 U^*_{b6}$ of \eq~(\ref{eq:corrseesaw}) remains constant when decreasing $\theta_3$. Therefore, the analysis performed here has extra degrees of freedom with respect to the adition of only 2 right-handed neutrinos considered in \Ref~\cite{Donini:2011jh} and, hence, obtaining the best fit values of \Refs~\cite{Kopp:2011qd,Giunti:2011gz} is easier. 

\section{Summary and Conclusions}
\label{sec:summary}

In this paper we have addressed the question of whether the best fit points found for short baseline neutrino oscillations trying to accommodate the MiniBooNE and LSND results of \Refs~\cite{Kopp:2011qd,Giunti:2011gz} can be reproduced with the more economical addition of 3 Majorana right-handed neutrinos to the SM. Such a scenario not only looks more natural from the point of view of the particle content, making the lepton and quark sectors symmetric and accommodating neutrino masses with a minimal extension, but also involves less degrees of freedom than the extension of the $3 \times 3$ neutrino mixing to $5 \times 5$. Indeed, the general $5 \times 5$ unitary mixing matrix of the 3+2 schemes contains 9 independent angles, 6 Dirac and 4 Majorana phases plus the 5 distinct neutrino masses, while the addition of 3 Majorana right-handed neutrinos only involves 9 independent angles, 4 Dirac and 2 Majorana phases and 3 masses (the other 3 being dependent combinations of the other masses and mixings).

In order to address this question we have developed a new parametrization of neutrino masses and mixings when the SM is extended by Majorana right-handed neutrinos. The advantage of this parametrization is that it is exact and, thus, valid for all regions of the parameter space. In particular, away from the seesaw limit that most existing parametrizations assume. This parametrization contains all the relations among masses and mixings defining a set of independent parameters from which all others can be easily derived. It can therefore be very useful for any scenario in which the seesaw relation, \ie, $M_N \gg m_D$, is not met with enough precision to justify the standard approximations as, for example, some cases of inverted or linear seesaw models. The parametrization is of course also valid for the standard seesaw and its correspondence to other popular parametrizations for this limit has been discussed.   

For the purpose at hand, the new parametrization allowed us to scan the parameter space down to very low Majorana masses, searching for points close to the best fit values of \Refs~\cite{Kopp:2011qd,Giunti:2011gz}. Four such points were chosen from a MCMC scan as samples to show that indeed the addition of 3 Majorana right-handed neutrinos can reproduce the more general 3+2 fits. These four points correspond to the two best fits available in the literature and the two mass hierarchies, normal and inverted, which currently can fit solar and atmospheric neutrino oscillations. All four samples are very close to the corresponding best fit regions showing that they can be accommodated for any choice of the mass hierarchy. Moreover, the mass of the third extra neutrino mass eigenstate, not necessary for the fit of short baseline neutrino data, can be chosen at any scale and still good fits are obtained. Thus, it could be used to try to account for the observed dark matter component of the Universe as a sterile neutrino candidate or made very heavy.

\begin{acknowledgments}
We acknowledge very useful discussions with Andrea Donini, Pilar Hernandez, Jacobo Lopez Pavon, Olga Mena, Silvia Pascoli and Michel Sorel. MB is supported by the European Union through the European Commission Marie Curie Actions Framework Programme 7 Intra-European Fellowship: Neutrino Evolution.
We also acknowledge support from the European Community under the European Commission Framework Programme 7 Design Study: EUROnu, Project Number 212372. 
\end{acknowledgments}

\end{document}